\documentclass[aps,twocolumn]{revtex4}
\usepackage{amsmath,amssymb}
\begin{document}   

\title{Holographic Scalar Fields Models of Dark Energy}

\author{Ahmad Sheykhi\footnote{sheykhi@uk.ac.ir}}

\address{Physics Department and Biruni Observatory, Shiraz University, Shiraz 71454,
         Iran\\
         Research Institute for Astronomy and Astrophysics of Maragha
         (RIAAM), P. O. Box 55134-441, Maragha, Iran  }

\begin{abstract}
Many theoretical attempts toward reconstructing the potential and
dynamics of the scalar fields have been done in the literature by
establishing a connection between holographic/agegraphic energy
density and a scalar field model of dark energy. However, in most of
these cases the analytical form of the potentials in terms of the
scalar field have not been reconstructed due to the complexity of
the equations involved. In this paper, by taking Hubble radius as
system's IR cutoff, we are able to reconstruct the analytical form
of the potentials as a function of scalar field, namely $V=V(\phi)$
as well as the dynamics of the scalar fields as a function of time,
namely $\phi=\phi(t)$, by establishing the correspondence between
holographic energy density and quintessence, tachyon, K-essence and
dilaton energy density in a flat FRW universe. The reconstructed
potentials are quite reasonable and have scaling solutions. Our
study further supports the viability of the holographic dark energy
model with Hubble radius as IR cutoff.

 \end{abstract}
\maketitle

\section{Introduction\label{Int}}
Holographic dark energy (HDE) models have got a lot of enthusiasm
recently, because they link the dark energy density to the cosmic
horizon, a global property of the universe, and have a close
relationship to the spacetime foam \cite{Ng1,pav2}. For a recent
review on different HDE models and their consistency check with
observational data see \cite{ser}. There are also a number of
theoretical motivations leading to the form of HDE, among which some
are motivated by holography and others from other principles of
physics. A fairly comprehensive motivations on HDE models can be
seen in \cite{li}. It is worthwhile to mention that in the
literature, various models of HDE have been investigated via
considering different system's IR cutoff. In the presence of
interaction between dark energy and dark matter, the simple choice
for IR cutoff could be the Hubble radius, $L=H^{-1}$ which can
simultaneously drive accelerated expansion and solve the coincidence
problem \cite{pav1,Zim}. Besides, it was argued that for an
accelerating universe inside the event horizon the generalized
second law does not satisfy, while the accelerating universe
enveloped by the Hubble horizon satisfies the generalized second law
\cite{Jia,shey1}. This implies that the event horizon in an
accelerating universe might not be a physical boundary from the
thermodynamical point of view. Thus, it looks that Hubble horizon is
a convenient horizon for which satisfies all of our accepted
principles in a flat Friedmann-Robertson-Walker (FRW) universe.

There has been a lot of interest in recent years in establishing a
connection between holographic/agegraphic energy density and scalar
field models of dark energy  (see e.g.
\cite{XZ,Setare,tachADE,quinADE}). These studies lead to reconstruct
the potential and the dynamics of the scalar fields according to the
evolution of the holographic/agegraphic energy density.
Unfortunately, in all of these cases
(\cite{XZ,Setare,tachADE,quinADE}) the analytical form of the
potentials have not been constructed as a function of scalar field,
namely $V=V(\phi)$, due to the complexity of the equations involved.
Recently, by implement a connection between scalar field dark energy
and HDE density and introducing a new IR cutoff, namely
$L^{-2}=\alpha H^2+\beta \dot{H}$, the authors of \cite{GO}
reconstructed explicitly the potentials and the dynamics of the
scalar fields, which describe accelerated expansion. Our work
differs from \cite{GO} in that we assume the pressureless dark
matter and HDE do not conserve separately but interact with each
other, while the authors of \cite{GO} have neglected the
contributions from dark matter and consequently no interaction
between two dark components. Besides, our system's IR cutoff differs
from that of Ref. \cite{GO}.

In this paper, by choosing Hubble radius $L=H^{-1}$ as system's IR
cutoff, we implement the connection between the HDE and scalar
fields models including quintessence, tachyon, K-essence and
dilaton energy density in a flat FRW universe. This simple and
most natural choice for IR cutoff allows us to reconstruct the
explicit form of potentials, $V=V(\phi)$, and also the dynamics of
the scalar fields as a function of time, namely $\phi=\phi(t)$.
\section{HDE with Hubble radius as an IR cut-off  \label{ORI}}
For the flat FRW universe, the first Friedmann equation is
\begin{eqnarray}\label{Fried}
H^2=\frac{1}{3M_p^2} \left( \rho_m+\rho_D \right),
\end{eqnarray}
where $\rho_m$ and $\rho_D$ are the energy density of dark matter
and dark energy, respectively. Taking the interaction between dark
matter and dark energy into account, the continuity equation maybe
written as  \cite{wang1,pav3}
\begin{eqnarray}
&&\dot{\rho}_m+3H\rho_m=Q, \label{consm}
\\&& \dot{\rho}_D+3H\rho_D(1+w_D)=-Q.\label{consq}
\end{eqnarray}
where $w_{D}=p_D/\rho_D$ is the equation of state (EoS) parameter
of HDE, and $Q$ stands for the interaction term. It should be
noted that the ideal interaction term must be motivated from the
theory of quantum gravity. In the absence of such a theory, we
rely on pure dimensional basis for choosing an interaction $Q$. It
is worth noting that the continuity equations imply that the
interaction term should be a function of a quantity with units of
inverse of time (a first and natural choice can be the Hubble
factor $H$) multiplied with the energy density. Therefore, the
interaction term could be in any of the following forms: (i)
$Q\propto H \rho_D$, (ii) $Q\propto H \rho_m$, or (iii) $Q\propto
H (\rho_m+\rho_D)$. We find out that for all three forms of $Q$,
the EoS parameter of HDE has a similar form, thus hereafter we
consider only the first case, namely $Q=3b^2 H \rho_D$, where
$b^2$ is a coupling constant.

We assume the HDE density has the form
\begin{equation}\label{rhoD}
 \rho_D=3c^2M_p^2H^2,
 \end{equation}
where $c^2$ is a constant and we have set the Hubble radius
$L=H^{-1}$ as system's IR cutoff. Inserting Eq. (\ref{rhoD}) in Eq.
(\ref{Fried}) immediately yields
\begin{equation}\label{u}
 u=\frac{1-c^2}{c^2}.
\end{equation}
where $u={\rho_m}/{\rho_D}$ is the energy density ratio. From Eq.
(\ref{u}) we see that the ratio of the energy densities is a
constant; thus the coincidence problem can be alleviated. It is
worth noting that in general the term $c^2$ in holographic energy
density can vary with time though very slowly \cite{pav4}. By slowly
varying we mean that $\dot{(c^2)}/c^2$ is upper bounded by the
Hubble expansion rate, $H$, i.e., \cite{pav4}
\begin{equation}\label{cdot}
 \frac{\dot{(c^2)}}{c^2}\leq H.
\end{equation}
Note that this condition must be fulfilled at all times; otherwise
the dark energy density would not even approximately be proportional
to $L^{-2}$, something at the core of holography \cite{pav4}. It was
argued that $c^2$ depends on the infrared length, $L$ \cite{pav4}.
For the case of $L=H^{-1}$, it was shown that one can take $c^2$
approximately constant in the late time where dark energy dominates
($\Omega_m<1/3$) \cite{pav4}. Since in the present work we study the
late time cosmology, and also for later convenience, we assume the
term $c^2$ to be a constant. Taking the time derivative of Eq.
(\ref{rhoD}), after using Friedmann equation (\ref{Fried}), we get
\begin{equation}\label{rhoD2}
 \dot{\rho}_D=-3c^2 H \rho_D(1+u+w_D).
 \end{equation}
Substituting Eq. (\ref{rhoD2}) in (\ref{consq}), after using
relations $Q=3b^2 H \rho_D$  and (\ref{u}), we obtain
\begin{equation}\label{wD}
 w_D=-\frac{b^2}{1-c^2}.
 \end{equation}
Therefore for constant parameters $c$ and $b$ the EoS parameter
becomes also a constant. In the absence of interaction, $b^2=0$, we
encounter dust with $w_D =0$. For the choice $L = H^{-1}$ an
interaction is the only way to have an EoS different from that for
dust \cite{pav1,Zim}. Let us note that in order to have $w_D < 0$ we
should have $c^2<1$. Besides, the acceleration expansion
($w_D<-1/3$) can be achieved provided $c^2>1-3b^2$. Thus this model
can describe the accelerated expansion if $1-3b^2<c^2<1$. Moreover,
$w_D$ can cross the phantom line ($w_D<-1$) provided $b^2>1-c^2$.
\section{Correspondence with scalar field models \label{SF}}
In this section we implement a correspondence between interacting
HDE and various scalar field models, by equating the equations of
state for this models with the equations of state parameter of
interacting HDE obtained in (\ref{wD}).
\subsection{Reconstructing holographic quintessence model}
In order to establish the correspondence between HDE and
quintessence scaler field,
 we assume the quintessence scalar field model of dark energy is
the effective underlying theory. The energy density and pressure of
the quintessence scalar field are given by \cite{cop}
\begin{eqnarray}\label{rhophi}
\rho_\phi=\frac{1}{2}\dot{\phi}^2+V(\phi),\\
p_\phi=\frac{1}{2}\dot{\phi}^2-V(\phi). \label{pphi}
\end{eqnarray}
Thus the potential and the kinetic energy term can be written as
\begin{eqnarray}\label{vphi}
&&V(\phi)=\frac{1-w_\phi}{2}\rho_{\phi},\\
&&\dot{\phi}^2=(1+w_\phi)\rho_\phi. \label{dotphi}
\end{eqnarray}
where $w_{\phi}=p_{\phi}/\rho_{\phi}$. In order to implement the
correspondence between HDE and quintessence scaler field, we
identify $\rho_\phi=\rho_D$ and $w_\phi=w_D$. Inserting Eqs.
(\ref{rhoD}) and (\ref{wD}) in (\ref{dotphi}) we reach
\begin{eqnarray}\label{dotphi1}
\dot{\phi}=\sqrt{3\left(1-\frac{b^2}{1-c^2}\right)} c M_p \frac
{\dot{a}}{a} .\label{dotphi2}
\end{eqnarray}
Integrating yields
\begin{eqnarray}\label{phi1}
\phi(a)=\sqrt{3\left(1-\frac{b^2}{1-c^2}\right)} c M_p \ln a,
\end{eqnarray}
where we have set $\phi(a_0=1)=0$ for simplicity.  Next we want to
obtain the scale factor as a function of $t$. Taking the time
derivative of Eq. (\ref{Fried}) and using (\ref{wD}) we find
\begin{eqnarray}\label{dotH}
\frac{\dot{H}}{H^2}=-\frac{3}{2}\left[1-\frac{b^2 c^2}{1-c^2}\right]
\end{eqnarray}
The fist integration gives
\begin{eqnarray}\label{dotH2}
H=\frac{da}{a dt}=\frac{2}{3\rm k t},
\end{eqnarray}
 where $\rm k= 1-\frac{b^2 c^2}{1-c^2}$. Integrating again we find
\begin{eqnarray}\label{at}
a(t)=t^{2/3\rm k}
\end{eqnarray}
Hence Eq. (\ref{phi1}) can be rewritten
\begin{eqnarray}\label{phi1t}
\phi(t)=\frac{2}{3 \rm k}c
M_p\sqrt{3\left(1-\frac{b^2}{1-c^2}\right)} \ln t.
\end{eqnarray}
Next we obtain the potential as a function of $\phi$. Combining Eq.
(\ref{wD}) with Eq. (\ref{vphi}) we reach
\begin{eqnarray}\label{vphi2}
V(\phi)=\frac{3}{2}c^2 M_p ^2 \left[1+\frac{b^2}{1-c^2}\right]H^2.
\end{eqnarray}
Using Eqs. (\ref{dotH2}) and (\ref{phi1t}) we obtain the explicit
expression for potential, namely
\begin{eqnarray}\label{vphi3}
V(\phi)&=&\frac{2 c^2M_p ^2}{3\rm k^2}\left[1+\frac{b^2}{1-c^2}\right]\times \\
\nonumber &&  \exp \left[ -3\frac{\rm
k}{cM_p}\left(3-\frac{3b^2}{1-c^2}\right)^{-1/2} \phi\right].
\end{eqnarray}
Let us discuss the condition for which the scale factor
(\ref{at}), and hence the obtained potential, leads to an
accelerated universe at the present time. Requiring $\ddot{a}>0$
for the present time, leads to $\rm k<2/3$ , which can be
translated into $c^2> (1+3b^2)^{-1}$. Note that the condition $\rm
k<2/3$ valid only for the late time where we have a dark energy
dominated universe. In general $\rm k$ depends on $c$, and for the
matter dominated epoch where $c$ is no longer a constant, then
$\rm k$ is also not a constant and varies with time. The obtained
exponential potential here is well-known in the literature for the
quintessence scalar field \cite{cop}. The cosmological dynamics of
this potential has been explored in detail \cite{cop}. In addition
to the fact that exponential potentials can give rise to an
accelerated expansion, they possess cosmological scaling solutions
\cite{cop,cop2,sahni} in which the field energy density
$\rho_{\phi}$ is proportional to the matter energy density
$\rho_{m}$.
\subsection{Reconstructing holographic tachyon model}
The tachyon field has been proposed as a possible candidate for dark
energy. A rolling tachyon has an interesting EoS whose parameter
smoothly interpolates between $-1$ and $0$ \cite{Gib1}. Thus,
tachyon can be realized as a suitable candidate for the inflation at
high energy \cite{Maz1} as well as a source of dark energy depending
on the form of the tachyon potential \cite{Padm}. Choosing different
self-interacting potentials in the tachyon field model lead to
different consequences for the resulting DE model. These give enough
motivations us to reconstruct tachyon potential $V(\phi)$ from HDE
model with Hubble radius as the IR cutoff. The correspondence
between tachyon field and various dark energy models such as HDE
\cite{Setare} and  agegraphic dark energy \cite{tachADE} has been
already established. The extension has also been  done to the
entropy corrected holographic and agegraphic dark energy models
\cite{ecde}. However, in all of these cases
\cite{Setare,tachADE,ecde} the explicit form of the tachyon
potential,$V=V(\phi)$, has not been reconstructed due to the
complexity of the equations involved.

The effective lagrangian for the tachyon field is given by
\cite{sen}
\begin{eqnarray}
 L=-V(\phi)\sqrt{1-g^{\mu\nu}\partial_\mu \phi \partial_\nu \phi},
 \end{eqnarray}
where $V(\phi)$ is the tachyon potential. The corresponding energy
momentum tensor for the tachyon field can be written in a perfect
fluid form
\begin{eqnarray}
 T_{\mu\nu}=(\rho+p)u_{\mu} u_\nu-p g_{\mu\nu},
 \end{eqnarray}
where $\rho$ and $p$ are, respectively, the energy density and
pressure of the tachyon and the velocity $u_\mu$ is
\begin{eqnarray}
 u_\mu=\frac{\partial_\mu \phi}{\sqrt{\partial_\nu \phi \partial^\nu
 \phi}}.
 \end{eqnarray}
The energy density and pressure of tachyon field are given by
\begin{equation}
\rho=-T_0^0=\frac{V(\phi)}{\sqrt{1-\dot\phi^2}},\label{rhoT}
\end{equation}
\begin{equation}
p=T_i^i=-V(\phi)\sqrt{1-\dot\phi^2}.
\end{equation}
Thus the EoS parameter of tachyon field is given by
\begin{equation}\label{wT}
w_T=\frac{p}{\rho}=\dot\phi^2-1.
\end{equation}
To establish the correspondence between HDE and tachyon field, we
equate  $w_D$ with $w_T$. From Eqs. (\ref{wD}) and (\ref{wT}) we
find
\begin{equation}\label{dotphiT}
\dot\phi^2=1-\frac{b^2}{1-c^2}.
\end{equation}
Integrating gives
\begin{equation}\label{phiT1}
\phi(t)=\left[1-\frac{b^2}{1-c^2}\right]^{1/2}t,
\end{equation}
where we set an integration constant to zero. Combining Eq.
(\ref{rhoT}) with (\ref{dotphiT}), the tachyon potential is obtained
as
\begin{eqnarray}\label{vphi1}
V(\phi)&=&3c^2 M_p^2 H^2 \frac{b}{\sqrt{1-c^2}},
\end{eqnarray}
Using Eqs. (\ref{dotH2}) and (\ref{phiT1}) we obtain tachyon
potential in terms of the scalar field
\begin{eqnarray}\label{vphi2}
V(\phi)&=&\frac{4c^2 M_p^2}{3\rm
k^2}\frac{b}{\sqrt{1-c^2}}\left(1-\frac{b^2}{1-c^2}\right) \frac
{1}{\phi^2} ,
\end{eqnarray}
From Eq. (\ref{phiT1}) we see that the evolution of the tachyon is
given by $\phi(t)\propto t$. The above inverse square power-law
potential corresponds to the one in the case of scaling solutions
\cite{cop,JM,EJ}.
\subsection{Reconstructing holographic K-essence model}
The scalar field model called K-essence is also employed to
explain the observed acceleration of the cosmic expansion. This
model is characterized by a scalar field with a non-canonical
kinetic energy. The most general scalar-field action which is a
function of $\phi$ and $X=-\dot{\phi}^2/2$ is given by \cite{arm}
\begin{equation}\label{kessaction}
    S=\int d^4x\sqrt{-g}P(\phi,X),
\end{equation}
where the lagrangian density $P(\phi,X)$ corresponds to a pressure
density. According to this lagrangian the energy density and the
pressure can be written as \cite{arm,cop}
\begin{eqnarray}\label{rhokess}
\rho(\phi,X)&=&f(\phi)(-X+3X^2),\label{rhok}\\
p(\phi,X)&=&f(\phi)(-X+X^2).\label{pk}
\end{eqnarray}
Therefore the EoS  parameter of the K-essence is given by
\begin{equation}\label{wk}
w_K=\frac{X-1}{3X-1}.
\end{equation}
Equating $w_K$ with the EoS parameter of HDE (\ref{wD}) one finds
\begin{equation}\label{X}
    X=\frac{1+b^2-c^2}{1+3b^2-c^2}.
\end{equation}
which implies that $X$ is  a positive constant ($c^2<1$). Indeed the
EoS parameter in Eq. (\ref{wk}) diverges for $X=1/3$. Let us
consider the cases with $X>1/3$ and $X<1/3$ separately. In the first
case where $X>1/3$, the condition $w_K < -1/3$ leads to $X<2/3$.
Thus we should have $1/3<X<2/3$ in this case. For example we obtain
the EoS of a cosmological constant ($w_K = -1$) for $X = 1/2$. In
the second case where $X<1/3$, we have $X-1<-2/3<0$, thus
$w_K=\frac{X-1}{3X-1}>0$. This means that we have no acceleration at
all. So this case is ruled out. As a result in K-essence model the
accelerated universe can be achieved provided $1/3<X<2/3$, which
translates into $1-3b^2<c^2<1$. This is consistent with our previous
discussions. Combining equation (\ref{X}) with $X=-\dot{\phi}^2/2$,
one gets
\begin{equation}\label{dotphik}
\dot{\phi}^2=\frac{2(1+b^2-c^2)}{1+3b^2-c^2},
\end{equation}
and thus we obtain the expression for the scalar field in the flat
FRW background
\begin{equation}\label{phik}
\phi(t)=\left[\frac{2(1+b^2-c^2)}{1+3b^2-c^2}\right]^{1/2}t ,
\end{equation}
where we have taken the integration constant $\phi_0$ equal to zero.
Taking the correspondence between HDE and K-essence into account,
namely $\rho_D=\rho(\phi,X)$, after using Eqs. (\ref{dotH2}) and
(\ref{phik}) we find
\begin{equation}\label{fphik}
f(\phi)=\frac{4c^2 M_p^2}{3\rm
k^2}\left[\frac{1+3b^2-c^2}{1-c^2}\right]\frac{1}{\phi^2}.
\end{equation}
Thus the K-essence potential $f(\phi)$ has a power law expansion.
From Eq. (\ref{phik}) we see that $\dot{\phi}=\rm const.$ This means
that the kinitic energy of K-essence becomes constant, though $\phi$
is not constant and evolves with time.
\subsection{Reconstructing holographic dilaton field}
The dilaton field may be used for explanation the dark energy puzzle
and avoids some quantum instabilities with respect to the phantom
field models of dark energy \cite{carroll}. The lagrangian density
of the dilatonic dark energy corresponds to the pressure density of
the scalar field has the following form \cite{tsuji}
\begin{equation}\label{dil1}
 p=-X+\alpha e^{\lambda\phi}X^2,
\end{equation}
where $\alpha$ and $\lambda$ are positive constants and
$X=\dot{\phi}^2/2$. Such a pressure (Lagrangian) leads to the
following energy density \cite{tsuji}
\begin{equation}\label{dil2}
\rho=-X+3\alpha e^{\lambda\phi}X^2.
\end{equation}
The EoS parameter of the dilaton dark energy can be written as
\begin{equation}\label{dil3}
w_d=\frac{1-\alpha e^{\lambda\phi}X}{1-3\alpha e^{\lambda\phi}X}.
\end{equation}
To establish the correspondence between HDE and dilaton field we
equate their EoS parameter, i.e. $w_d=w_D$. We reach
\begin{equation}\label{dil4}
\frac{1-\alpha e^{\lambda\phi}X}{1-3\alpha
e^{\lambda\phi}X}=-\frac{b^2}{1-c^2}.
\end{equation}
Using relation $X=\dot{\phi}^2/2$, and integrating with respect to
$t$ we find
\begin{equation}\label{dil5}
\phi=\frac{2}{\lambda}\ln
\left[\frac{\lambda}{\sqrt{2\alpha}}\left(\frac{1+b^2-c^2}{1+3b^2-c^2}\right)^{1/2}t\right]
\end{equation}
The existence of scaling solutions for the dilaton was studied in
\cite{tsuji} and was found that in this case the scaling solution
corresponds to $X e^{\lambda \phi}=\rm const.$, which has the
solution $\phi(t)\propto \ln t$. The results we found here by
equating the EoS parameter of HDE and dilaton field are consistent
with those obtained in \cite{tsuji}.
\section{Conclusion and discussion\label{CONC}}
In this paper by choosing the Hubble radius as system's IR cutoff
for interacting HDE model, we established a connection between the
scalar field model of dark energy including quintessence, tachyon,
K-essence and dilaton energy density and holographic energy
density. As a result, we reconstructed the analytical form of
potentials namely $V=V(\phi)$ as well as the dynamics of the
scalar fields as a function of time explicitly, namely
$\phi=\phi(t)$ according to the evolutionary behavior of the
interacting HDE model. The obtained expressions for the potentials
are quite reasonable and lead to scaling solutions. Our studies
favor the $L=H^{-1}$ IR cutoff as a viable phenomenological model
of HDE.

Finally, I would like to mention that usually, for the sake of
simplicity, the term $c^2$ in holographic energy density
(\ref{rhoD}) is assumed constant. However, one should bear in mind
that it is more general to consider it a slowly varying function of
time, $c^2(t)$ \cite{pav4}. In this case the EoS parameter given in
(\ref{wD}) is no longer a constant. As a result we cannot integrate
easily the resulting equations in section III and find the
analytical form of the potentials. It is important to note that,
although, with implement the correspondence between this HDE model
and scalar field models, the EoS of scalar fields are assumed to be
fixed, nevertheless, neither $\phi$ nor $V(\phi)$ are not constant
and they still evolve with time. In the present work for simplicity
we have taken $c=\rm const.$ The correspondence between HDE and
scalar field models with varying $c^2$ term is under investigation
and will be addressed elsewhere.

\acknowledgments{I am grateful to the referee for valuable
comments and suggestions, which have allowed me to improve this
paper significantly. I sincerely thank Prof. Diego Pavon for
constructive comments on an earlier draft of this paper. Special
thanks go to Dr. E. Ebrahimi for many helpful discussions. This
work has been supported financially by Research Institute for
Astronomy and Astrophysics of Maragha (RIAAM) under research
project No. 1/2030.}

\end{document}